\documentclass[acus]{JAC2000}

\usepackage{graphicx}

\begin{document}
\title{
Emittance Growth from the Thermalization of Space-Charge 
Nonuniformities\thanks{This work was 
performed under the auspices of the U.S. Department of 
Energy by University of California at Lawrence Livermore National 
Laboratory and Lawrence Berkeley National Laboratory under contract 
Nos. W-7405-Eng-48 and DE-AC03-76SF00098.}}

\author{Steven M. Lund, John J. Barnard, and Edward P. Lee \\  
Lawrence Livermore and Berkeley National Laboratories, University of 
California, USA}

\maketitle

\baselineskip=11.6pt 

\begin{abstract} 
Beams injected into a linear focusing channel 
typically have some degree of space-charge 
nonuniformity.  In general, injected particle distributions with 
systematic charge nonuniformities are not  
equilibria of the focusing channel and launch a  
broad spectrum of collective modes.  These  
modes can phase-mix and have nonlinear wave-wave 
interactions which, at high space-charge intensities, results 
in a relaxation to a more thermal-like distribution characterized 
by a uniform density profile.  This thermalization can   
transfer self-field energy from the initial 
space-charge nonuniformity to the local particle 
temperature, thereby increasing beam phase space 
area (emittance growth).  In this paper, we 
employ a simple kinetic model of a continuous focusing 
channel and build on previous work that applied system 
energy and charge 
conservation\cite{Reiser-94,Davidson-90} to 
quantify emittance growth associated with 
the collective thermalization of an initial azimuthally 
symmetric, rms matched beam with a radial density profile 
that is hollowed or peaked.  This emittance growth is shown to 
be surprisingly modest even for high beam intensities with  
significant radial structure in the initial density profile.  
%
\end{abstract}

\vspace{-0.2 true cm}
\section{INTRODUCTION}
\vspace{-0.1 true cm}

Experiments with high-current, heavy-ion 
injectors have observed significant space-charge nonuniformities 
emerging from the source. 
Sharp density peaks on the  
radial edge of beam have been measured,  
but the local incoherent thermal spread of particle velocities 
(i.e., the particle temperature) across the beam is anticipated 
to be fairly uniform since the beam is emitted from 
a constant temperature surface.  When such a distribution 
is injected into a linear transport channel, it  
will be far from an equilibrium condition (i.e., particles out  
of local radial force balance), and a broad spectrum 
of collective modes will be launched.  

The spatial average particle temperature of a heavy ion beam 
emerging from an injector is typically measured as several times 
what one would infer from the source thermal temperature 
($\sim 0.1$eV) and 
subsequent beam envelope compressions, with $\bar{T}_x \sim 20$eV  
where $\bar{T}_x \sim [\epsilon_x^2/(2R^2)]{\cal E}_b$.  On 
the other hand, the radial change in potential energy 
from beam center to 
edge is $q\Delta\phi \sim 2.25$keV for a beam with line-charge 
density $\lambda \sim 0.25\mu$C/m 
($\Delta\phi \sim \lambda/(4\pi\epsilon_0)$).  If even a 
small fraction of such space-charge energy is thermalized 
during collective relaxation, large temperature and 
emittance increases can result.  

In this paper, we employ conservation constraints 
to better estimate emittance increases from 
collective thermalization of normal mode perturbations 
resulting from initial space-charge nonuniformities 
characteristic of intense beam injectors.  Past 
studies have employed analogous techniques to 
estimate emittance increases resulting from the thermalization 
of initial rms mismatches in the beam envelope and space-charge 
nonuniformities associated with combining 
multiple beams and other 
processes\cite{Reiser-94,Davidson-90,Wangler-85}.  


\vspace{-0.2 true cm}
\section{THEORETICAL MODEL}
\vspace{-0.1 true cm}

We analyze an infinitely long, unbunched ($\partial/\partial z = 0$) 
nonrelativistic beam composed of a single species of particles of mass 
$m$ and charge $q$ propagating with constant axial 
kinetic energy ${\cal E}_b$.  Continuous radial focusing is provided 
by an external force that is proportional to the 
transverse coordinate ${\bf x}$, i.e., 
${\bf F}_{\rm ext} = -2{\cal E}_b k_{\beta 0}^2 {\bf x}$, where  
$k_{\beta 0} =\;$const is the betatron wavenumber of particle oscillations 
in the applied focusing field.  
For simplicity, we neglect particle collisions and correlation effects, 
self-magnetic fields, and employ an electrostatic model 
and describe the transverse evolution of the beam as a function of axial 
propagation distance $s$ in terms of a single-particle 
distribution function $f$ that is a function of $s$, 
and the transverse position ${\bf x}$ and angle ${\bf x}' = d{\bf x}/ds$ of 
a single particle.  This evolution is described by 
the Vlasov equation\cite{Davidson-90}, 
\vspace{-0.2 true cm}
\begin{eqnarray}
\left\{ {\partial \over \partial s} + {\partial H \over \partial{\bf x}'}\cdot 
{\partial \over \partial{\bf x}} - {\partial H \over \partial{\bf x}}\cdot 
{\partial \over \partial{\bf x}'} \right\} f({\bf x},{\bf x}',s) = 0,
\label{Vlasov-eqn} 
\end{eqnarray}
where 
$H = {\bf x}'^2 /2  + k_{\beta 0}^2 {\bf x}^2 /2  + (q/2{\cal E}_b )\phi$ 
%
is the single-particle Hamiltonian and the self-field 
potential $\phi$ satisfies the Poisson equation 
(CGS units here and henceforth)  
\vspace{-0.2 true cm}
\begin{eqnarray}
\nabla^2\phi = -4\pi q\int\! d^2x'\;  f
\label{Poisson-eqn}  
\end{eqnarray}
subject to the boundary condition $\phi(r=r_p) = 0$ at the conducting 
pipe radius $r = |{\bf x}| = r_p = \;$const.  


If no particles are lost in the beam evolution, the 
Vlasov-Poisson system possesses global constraints 
corresponding to the conservation of system 
charge ($\lambda$) and scaled energy ($U$) per unit axial 
length, 
\vspace{-0.2 true cm}
\begin{eqnarray}
\lambda &=& q\int\! d^{2}x \!\!\int\! d^{2}x' \; f \;\; = {\rm const},  
\nonumber \\  
U &=& {1 \over 2}\langle {\bf x}'^2 \rangle  +   
      {k_{\beta 0}^2 \over 2} \langle {\bf x}^2 \rangle   
   +  {q \over 2{\cal E}_b\lambda } W \; = \; {\rm const}.
\label{cons-constr} 
\end{eqnarray}
Here, $W \equiv \int\! d^{2}x \; |\nabla\phi |^2 /( 8\pi)$ 
is the self-field energy of the beam per unit axial length and 
$\langle \xi \rangle \equiv (\int\! d^{2}x \int\! d^{2}x' \; 
\xi\; f  \;)/(\int\! d^{2}x \int\! d^{2}x'\; f\; )$ is 
a transverse statistical average of 
$\xi$ over the beam distribution $f$.  Note that 
$U$ includes both particle kinetic energy and the field 
energy of the applied and self-fields.   These conservation laws  
follow directly from Eqs.\ (\ref{Vlasov-eqn})-(\ref{Poisson-eqn}) and   
provide powerful constraints on the nonlinear evolution of the system.  

Moment descriptions of the beam provide a simplified understanding 
of beam transport.  For an azimuthally symmetric 
beam ($\partial/\partial\theta = 0$), 
a statistical measure of the beam edge radius 
$R \equiv 2\langle x^2 \rangle^{1/2}$ is employed.  
Note that $R$ is the edge radius of a beam with 
uniformly distributed space-charge.  Any 
axisymmetric solution 
to the Vlasov-Possion system will be 
consistent with the rms envelope 
equation\cite{Reiser-94}
\vspace{-0.2 true cm}
\begin{eqnarray}
{d^2 R \over ds^2} + k_{\beta 0}^2 R - {Q \over R} - 
{\epsilon_x^2  \over R^3} = 0 \; .
\label{envelope-eqn}  
\end{eqnarray}
Here, $Q = q\lambda/{\cal E}_b = \;$const is the self-field perveance and 
$\epsilon_x = 4[ \langle x^2 \rangle\langle x'^2 \rangle - 
\langle x x' \rangle^2 ]^{1/2}$ 
%
%
is an edge measure of the rms $x$-emittance of the beam and is a 
statistical measure of the beam area in $x$-$x'$ 
phase-space (i.e., beam quality).  For general distributions, 
$\epsilon_x$ is not constant and evolves  
according to the full Vlasov-Poisson system.  


\vspace{-0.2 true cm}
\section{NONUNIFORM DENSITY PROFILE}
\vspace{-0.1 true cm}

\begin{figure}[htb]
\centering
\includegraphics*[width=60mm]{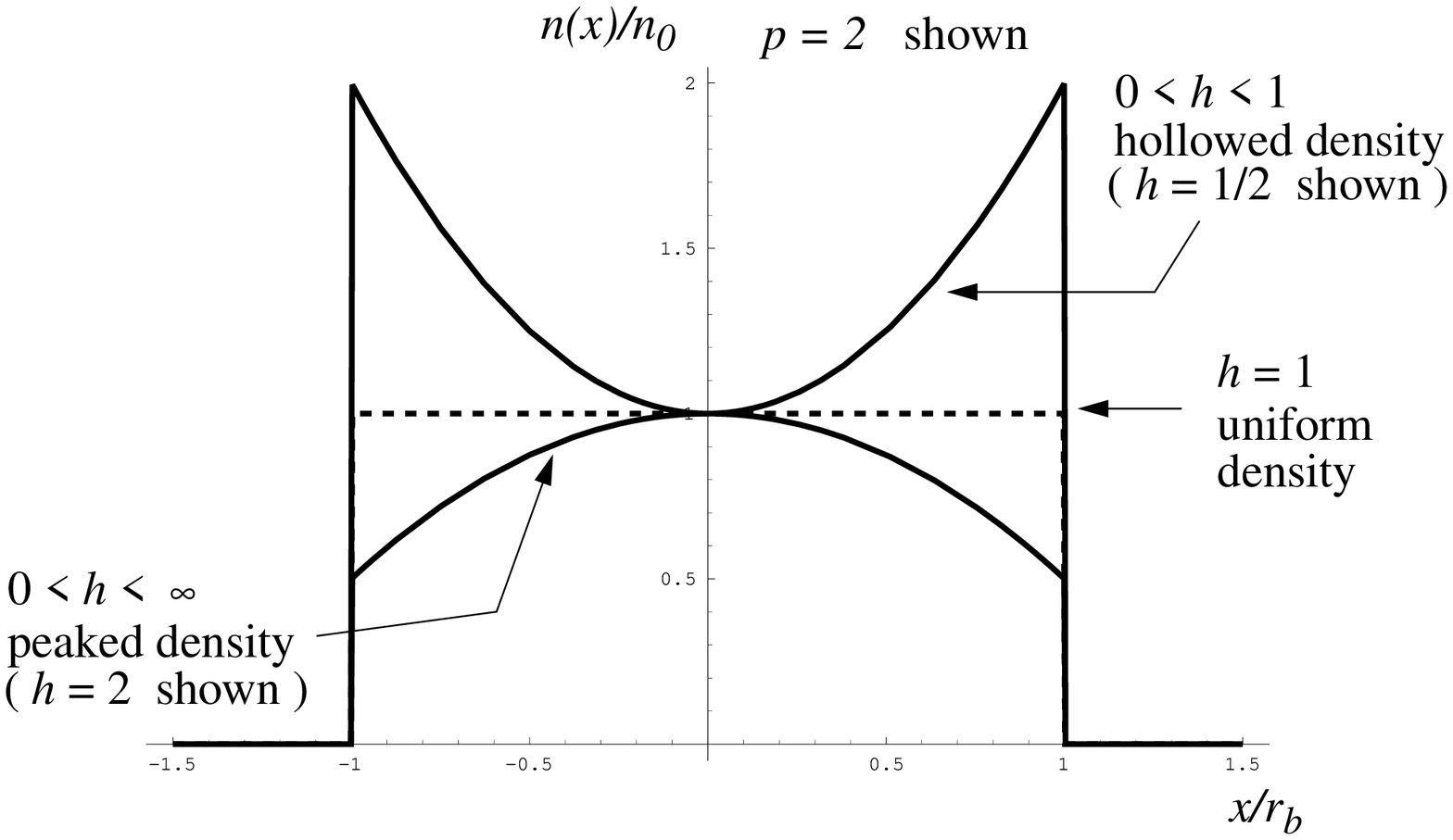}
\vspace{-0.4 true cm}
\caption{Uniform, hollowed, and peaked density profiles.}
\vspace{-0.2 true cm}
\label{Fig-dens-hollowed}
\end{figure}

We examine an beam with an azimuthally symmetric 
radial density profile $n = \int\! d^2x'\; f$ given by 
\vspace{-0.2 true cm}
\begin{eqnarray}
n(r) = 
\left\{\begin{array}{ll} 
n_0\left[ 1 - {1-h \over h}\left( {r \over r_b} \right)^p \right],  
         & \mbox{$0 \leq r \leq r_b$,} \\  
0,                                                                  
         & \mbox{$r_b < r  \leq r_p$.} 
\end{array}\right. 
\label{dens-hollowed}
\end{eqnarray}
Here, $r_b$ is the physical edge-radius of the 
beam, $n_0 = n(r=0)$ is the on-axis ($r=0$) 
beam density, and $h$  and $p$ are ``hollowing'' 
[$0 \leq h \leq \infty$, $h = n(r=r_b)/n(r=0)$, 
$p \geq 0$] and radial steepening parameters 
associated with the density nonuniformity.  
This density profile is illustrated in 
Fig.\ \ref{Fig-dens-hollowed} for the steepening index  
$p = 2$ and hollowing factors $h = 1$ (uniform), 
$h = 1/2$ (hollowed), and $h = 2$ (peaked).  
The hollowing parameter $h$ has range $0 \leq h < 1$ 
for an on-axis hollowed beam and $0 \leq 1/h < 1$ for an 
on-axis peaked beam.  The limit $h \rightarrow 1$ corresponds 
to a uniform density beam and 
$h,\; 1/h \rightarrow 0$ 
correspond 
to hollowed 
and peaked beams with the density approaching zero 
on-axis and at the beam edge ($r = r_b$), 
respectively.  For large steepening index $p \gg 1$, 
the density gradient will be significant only near the 
radial edge of the beam ($r \simeq r_b$), and the 
density is uniform for $h = 1$ regardless of $p$.    

The beam line-charge density ($\lambda$) and rms 
edge-radius ($R$) are related to the 
parameters in Eq.\ (\ref{dens-hollowed}) by 
\vspace{-0.2 true cm}
\begin{eqnarray}
\lambda = \int\! d^2x\; n \; = \pi q n_0 r_b^2 \left[ {(ph+2) \over (p+2)h} 
\right] 
\nonumber \\ 
R = 2\langle x^2 \rangle^{1/2} = 
\sqrt{ {(p+2)(ph+4) \over (p+4)(ph+2)} } r_b 
\label{dens-hollowed-params} 
\end{eqnarray}
%
%
Using these expressions, the Poisson equation 
(\ref{Poisson-eqn}) can be solved for the potential $\phi$ 
corresponding to the density profile (\ref{dens-hollowed}) and 
used to calculate the self-field energy $W$ as 
\vspace{-0.2 true cm}
\begin{eqnarray}
W = \lambda^2\left\{ 
{1 \over (ph+2)^2}\left[ {(p+2)^2 h^2 \over 4} + 
{2(1-h)^2 \over p+2} + 
\right.\right. 
\nonumber \\  
 \left.\left. 
{4(p+2)h(1-h) \over p+4} \right] + 
\ln\left[ \sqrt{(p+2)(ph+4) \over (p+4)(ph+2)}{r_p \over R} \right] \right\} .
\label{dens-hollowed-field-eng}
\end{eqnarray}
%

It is convenient to {\em define} an average phase advance 
parameter $\sigma$ for the density profile (\ref{dens-hollowed}) in 
terms of an envelope matched ($R' = 0 = R''$), rms equivalent beam with 
uniform density ($h=1$) and the same perveance ($Q$) and emittance 
($\epsilon_x$) as the (possibly mismatched) beam with a  
nonuniform density profile ($h \neq 1$).  Denoting the phase advance 
per unit axial length of transverse 
particle oscillations in the matched equivalent 
beam in the presence and absence of space-charge by $\sigma$ and 
$\sigma_0$, we adapt a normalized space-charge parameter 
$\sigma/\sigma_0 \equiv \sqrt{k_{\beta 0}^2 - Q/R^2}/k_{\beta 0}$.   
%
%
The limits $\sigma/\sigma_0 \rightarrow 0$ and 
$\sigma/\sigma_0 \rightarrow 1$ 
correspond to a cold, space-charge dominated beam and a warm, 
kinetic dominated beam, respectively.  
%
%
Note that this measure applies only 
in an equivalent beam sense.  In general, distributions $f$ consistent 
with the density profile (\ref{dens-hollowed}) will not be 
equilibria ($d/ds \neq 0$) of the 
transport channel and will evolve leaving $\sigma$ ill defined.  

\vspace{-0.2 true cm}
\section{EMITTANCE GROWTH}
\vspace{-0.1 true cm}

We consider an initial beam distribution $f$ with a density profile 
given by Eq.\ (\ref{dens-hollowed}) and an {\em arbitrary} 
``momentum'' distribution in ${\bf x}'$.  Such an initial distribution 
is not, in general, an equilibrium of the focusing channel and a 
spectrum of collective modes will be launched (depending on the full 
initial phase-space structure of $f$).  These modes 
will phase-mix, have nonlinear wave-wave 
interactions, etc., driving relaxation processes that have been 
observed in PIC simulations to  
cause the beam space-charge distribution to become more 
uniform for the case of high beam intensities.  The 
conservation constraints (\ref{cons-constr}) are employed to connect the 
parameters of an initial (subscript $i$), nonuniform density beam 
with $h \neq 0$ with those of a final (subscript $f$), 
azimuthally symmetric and rms envelope 
matched beam ($R_f' = 0 = R_f''$) with uniform density ($h = 1$).   

Employing Eqs.\ (\ref{envelope-eqn})-(\ref{dens-hollowed-field-eng}),  
conservation of charge 
($\lambda_i = \lambda_f \equiv \lambda$) 
and system energy ($U_i = U_f$) can be combined into an single 
equation of constraint expressible as 
\vspace{-0.2 true cm}
\begin{eqnarray}
{ (R_f/R_i)^2 -1 \over 1 - (\sigma_i/\sigma_0)^2 } + 
{ p(1-h)[4 + p + (3+p)h] \over (p+2)(p+4)(2+ph)^2 } 
\nonumber \\ 
-\ln\left[\sqrt{(p+2)(ph+4) \over (p+4)(ph+2)}{R_f \over R_i}\right]
= {{\cal E}_b \over 2q\lambda}(R_i R_i')'
\label{f-i-rms-constr}  
\end{eqnarray}
Here, $h$ and $p$ are the hollowing factor and index of the initial 
density profile, $\sigma_i/\sigma_0$ is the 
initial space-charge intensity,  
and $[{\cal E}_b/(2q\lambda)](R_i R_i')'$ is a parameter that measures 
the initial envelope mismatch of the beam.  This nonlinear constraint 
equation can be solved numerically for fixed 
$h$, $p$, $\sigma_i/\sigma_0$ and $[{\cal E}_b/(2q\lambda)](R_i R_i')'$ 
to determine the ratio of final to initial rms radius of 
the beam ($R_f/R_i$).  
%
%
Employing the envelope equation (\ref{envelope-eqn}), 
the ratio of final to initial beam emittance is 
expressible as 
%
%
\vspace{-0.2 true cm}
\begin{eqnarray}
{\epsilon_{xf} \over \epsilon_{xi}} = {R_f \over R_i} 
\sqrt{ (R_f/R_i)^2 - [1-(\sigma_i/\sigma_0)^2] \over 
       (\sigma_i/\sigma_0)^2 - R_i''/(k_{\beta 0}^2 R_i) } .  
\label{f-i-emit-constr}
\end{eqnarray}
%
Eqs.\  (\ref{f-i-rms-constr}) and (\ref{f-i-emit-constr}) allow 
analysis of emittance growth from the thermalization 
of initial space-charge nonuniformities. 

\begin{figure}[htb]
\centering
\includegraphics*[width=85mm]{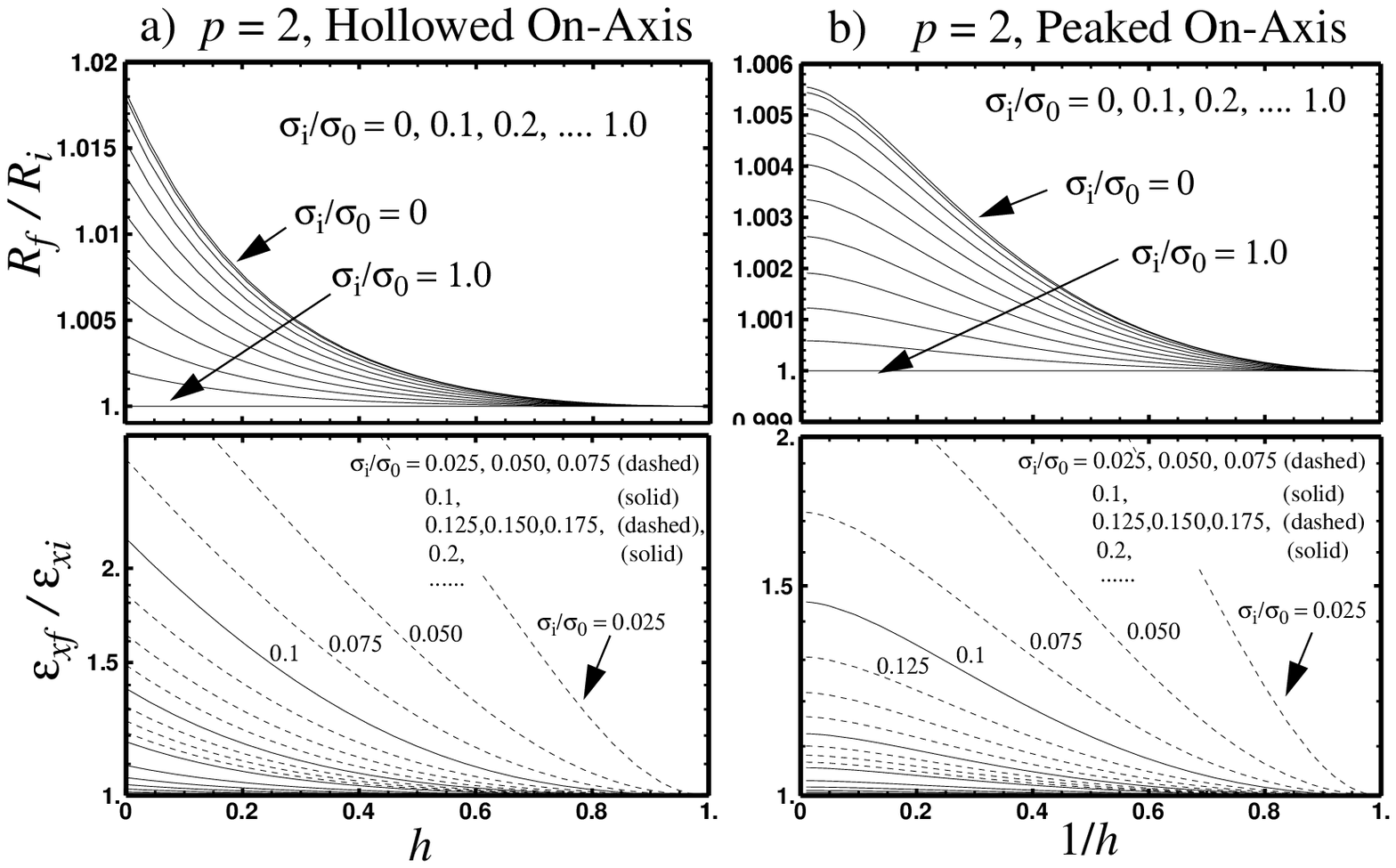}
\vspace{-0.6 true cm}
\caption{
Ratio of final to initial rms beam size 
($R_f/R_i$) and emittance ($\epsilon_{xf}/\epsilon_{xi}$) 
%
%
verses $h$ (a, hollowed beam) and $1/h$ (b, peaked beam).  
%
}
\vspace{-0.2 true cm}
\label{Fig-growth1}
\end{figure}

We numerically solve 
Eqs.\ (\ref{f-i-rms-constr}) and (\ref{f-i-emit-constr}) to 
plot (Fig.\ \ref{Fig-growth1}) the growth in rms beam 
radius ($R_f/R_i$) and emittance 
($\epsilon_{xf}/\epsilon_{xi}$) due to the relaxation of 
an initial rms matched beam 
($R_i' = 0 = R_i''$)
with nonuniform hollowed 
and peaked density profiles to a final uniform, 
matched profile.  Final to initial beam ratios
are shown for hollowing index of $p=2$ and are plotted 
verses the ``hollowing factors'' $h$ (hollow initial density) 
and $1/h$ (peaked initial density) for families of 
$\sigma_i/\sigma_0$ ranging from 
$\sigma_i/\sigma_0 \rightarrow 0$ to $\sigma_i/\sigma_0 \rightarrow 1$.  
Growths are larger for the initially hollowed profile 
than the peaked profile and increase with stronger 
space-charge (smaller $\sigma_i/\sigma_0$).  However, the 
change in rms radius ($R_f/R_i$) is small in all cases, even 
for strong space-charge with strong hollowing ($h \rightarrow 0$) and 
peaking ($1/h \rightarrow 0$) parameters.  Moreover, the increases in 
beam emittance ($\epsilon_{xf}/\epsilon_{xi}$) are surprisingly modest 
(factor of 2 and less) 
for intense beam parameters with $\sigma_i/\sigma_0 \sim 0.1$ and greater.    
At fixed $\sigma_i/\sigma_0$ and increasing steeping factor $p$, 
similar modest growth factors are seen for hollowed beams for  
all but the most extreme hollowing factors ($h \sim 0.1$ and less), 
and as expected, much less growth is seen for peaked 
beams (closer to uniform).  

\vspace{-0.2 true cm}
\section{DISCUSSION}
\vspace{-0.1 true cm}

\begin{figure}[htb]
\centering
\includegraphics*[width=50mm]{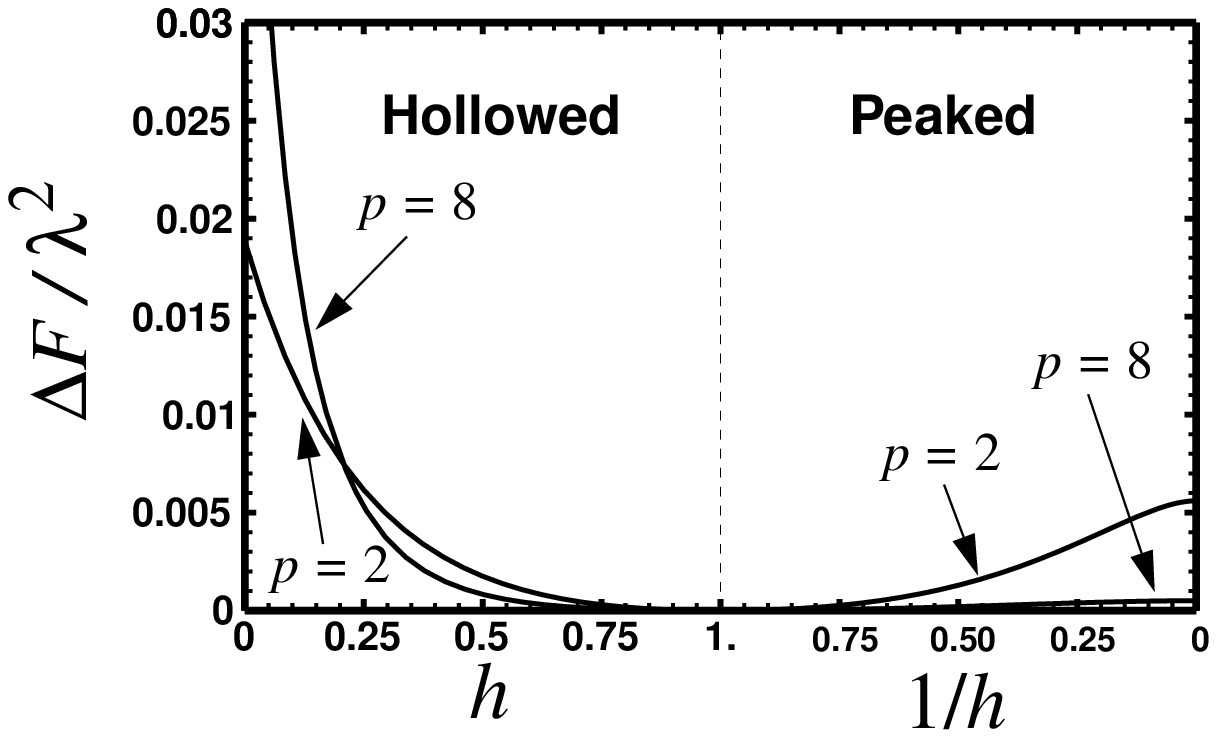}
\vspace{-0.4 true cm}
\caption{Free energy verses hollowing factors $h$ and $1/h$.}
\vspace{-0.2 true cm}
\label{Fig-free-eng}
\end{figure}

The modest emittance growth at high beam intensities 
can be understood as general 
beyond the specific model employed.  
Even for significant increases in emittance 
$\epsilon_x$, the rms matched beam size is 
given to a good approximation 
by the envelope equation (\ref{envelope-eqn}) with  
the emittance term neglected.  In this case, 
$R \simeq \sqrt{Q}/k_{\beta 0} =$const during the 
beam evolution and hence $R_f \simeq R_i$.   
Employing the method of Lagrange multipliers, the free 
electrostatic energy of the system at fixed rms radius   
($R$) and line-charge ($\lambda$) can be expressed as 
$F = W - \int\! d^2x\; (\mu_1 r^2  + \mu_2 )n$ with 
$\mu_{1,2} =$const. Taking variations $\delta\phi$ subject 
to the Poisson equation (\ref{Poisson-eqn}), one 
obtains to arbitrary order in $\delta\phi$,
\vspace{-0.2 true cm}
\begin{eqnarray} 
\delta F = \int\! d^2x\; (q\phi - \mu_1 {\bf x}^2 - \mu_2)\delta n \; + \;  
\int\! d^2x\; {|\nabla\delta\phi |^2 \over 8\pi} .
\end{eqnarray}
Thus, constrained extrema of $F$ satisfy 
$q\phi = \mu_1 {\bf x}^2 + \mu_2$, corresponding to a 
uniform density beam centered on-axis.  Variations about this 
extremum satisfy $\delta F > 0$ and are second order 
in $\delta\phi$.  Thus, the available electrostatic energy 
for thermalization induced emittance increase is modest for 
any smooth density profile.  This can be demonstrated for our 
specific example using equation (\ref{dens-hollowed-field-eng}) 
to plot $\Delta F = W_i - W_f$ with $R_i = R_f$ verses $h$ and $1/h$
for $p = 2,8$ (Fig.\ \ref{Fig-free-eng}).  

It has been shown that the rms beam size and emittance undergo 
very small {\em decreases} on relaxation from a uniform density 
beam to thermal equilibrium 
over the full range of $\sigma_i/\sigma_0$ ( 
${\rm Min}[\epsilon_{xf}/\epsilon_{xi}] \simeq 0.97$ at 
$\sigma_i/\sigma_0 \simeq 0.45$)\cite{Lund-95}.  Thus if one 
views the relaxation as a multi-step procedure using the 
conservation constraints to connect the initial nonuniform 
profile to a uniform profile and then a thermal profile, 
any emittance growth will be experienced in the first step.  
This result together with the variational argument above 
show that the emittance growth results 
presented should be relatively insensitive to the form 
of the final distribution.  

Finally, caveats should be given for validity of the theory.  
First, the model assumes no generation of halo in the final 
state and that the initial nonuniform beam can be perfectly 
rms envelope matched.  Initial mismatches can lead to halo 
production and provide a large source of free energy which, 
if thermalized, can lead to significant emittance 
growth\cite{Reiser-94}.  Also, although the velocity space 
distribution is arbitrary in the present theory, choices 
that project onto broader spectrums of modes will more 
rapidly phase mix and thermalize.  
Small applied nonlinear fields tend to 
enhance this relaxation rate.  Initial simulation 
results in a full AG lattice are 
consistent with model predictions presented here and will 
be presented in future work.  

%
%
 
%
%

\vspace{-0.2 true cm}

\end{document}